\documentclass{elsart}

\usepackage{amssymb}

\usepackage{graphicx}
\usepackage{epsfig,latexsym}

\newcommand{\be}{\begin{eqnarray}}
\newcommand{\ee}{\end{eqnarray}}

\begin{document}

\begin{frontmatter}



\title{Ground State Quarkonium Spectral Functions Above Deconfinement}


\author[label1]{\'Agnes M\'ocsy}
\author[label2]{, P\'eter Petreczky}
\author[label3]{, Jorge Casalderrey-Solana}

\address[label1]{RIKEN-BNL Research Center, Brookhaven National Laboratory, Upton, NY 11973, USA}

\address[label2]{Physics Department, Brookhaven National Laboratory, Upton, NY 11973, USA}

\address[label3]{Nuclear Science Division, MS 70R319, Lawrence Berkeley National Laboratory, Berkeley, CA 94720 }

\begin{abstract}
We discuss the temperature-dependence of S-wave quarkonium spectral functions in a nonrelativistic Green's function approach and compare these to lattice QCD results. 
\end{abstract}

\end{frontmatter}

For the last twenty years the melting of heavy quark bound states has been considered an unambiguous signal for deconfinement \cite{Matsui:1986dk}. The idea is that in the deconfined phase of QCD the color force between a heavy quark and antiquark is  Debye-screened. When the temperature-dependent range of screening is smaller than the size of the boundstate then this will dissociate. Understanding the modification of the properties of heavy quarkonia in a hot medium is therefore essential for understanding deconfinement. Phenomenological studies based on potential models have been in the recent years accompanied by lattice QCD calculations. 

The temperature-dependence of the meson correlators  can provide information about the fate of quarkonia states in a hot medium above deconfinement. The Euclidean correlation functions of meson currents $G(\tau,T)$ are reliably  calculated on the lattice. Their  spectral representation  
$ G(\tau,T)=\int d\omega \sigma(\omega,T)K(\omega,\tau,T)$
allows for the extraction of the spectral functions $\sigma(\omega,T)$ using the Maximum Entropy Method \cite{Umeda:2002vr}.  
Spectral functions obtained in the pseudoscalar channel from quenched lattice calculations suggest little modification of the ground states above $T_c$: The 1S charmonium state $\eta_c$ survives up to $1.5T_c$ and the 1S bottomonium  $\eta_b$ shows very small change up to $2.3T_c$.  Furthermore, no thermal shift in the ground state masses has been observed, and within errors the amplitudes are also  unchanged \cite{Umeda:2002vr}.  

These lattice results are in contradiction with the original potential model predictions. Potential models assume that the interaction between a heavy quark and antiquark is instantaneous and is mediated by a potential. The properties of a bound state are determined by solving the Schr\"odinger equation. At zero temperature the Cornell potential seems to describe quarkonia spectroscopy acceptably well. At finite temperature the form of the potential is not known. It is even questionable whether a temperature-dependent quark-antiquark potential is adequate for the understanding of the properties of quarkonia at finite temperature. Nevertheless, in lack of other theoretical means potential models have been widely used to investigate quarkonium states at finite temperature. The expectation was that the $J/\psi$ (and $\eta_c$) will dissociate by  $1.1T_c$. 

After the appearance of the lattice data potential models have been reconsidered using different temperature dependent screened potentials. The internal energy of a heavy quark-antiquark pair as determined on the lattice \cite{Kaczmarek:2003dp} became a popular choice for the potential \cite{Shuryak:2003ty} . A combination of the lattice internal and free energy has also been suggested \cite{Wong:2004zr}. Common in these potentials is that they contain temperature-dependent screening, and yield quarkonium dissociation temperatures in accordance with the lattice data. 

The first study of quarkonium correlators using potential models \cite{Mocsy:2004bv} points out that even though certain screened potentials can reproduce qualitative features of the lattice spectral function, such as the survival of the 1S state well above $T_c$, the temperature dependence of neither the correlators, nor the properties of the states are reproduced. These results are obtained using a spectral function, designed as the sum of bound state/resonance contributions and the perturbative continuum above a threshold. This Ansatz for the spectral function is justified when there is a sizable gap between the resonance and the continuum. As the temperature increases the gap becomes smaller and the Ansatz might not be the best. To avoid the artificial separations of bound state contributions and the continuum to the spectral function we calculate the full non-relativistic Green's function \cite{Mocsy:2006zd,{new}}. The imaginary part of the Green's function yields the quarkonium spectral function. Such method was used to determine properties of topomonium states \cite{Strassler:1990nw}, and then of binary light quark states \cite{Casalderrey-Solana:2004dc}. The main advantage of this approach is that resonances and continuum are considered together. 

\begin{figure}[htbp]
\begin{minipage}[htbp]{5cm}
\epsfig{file=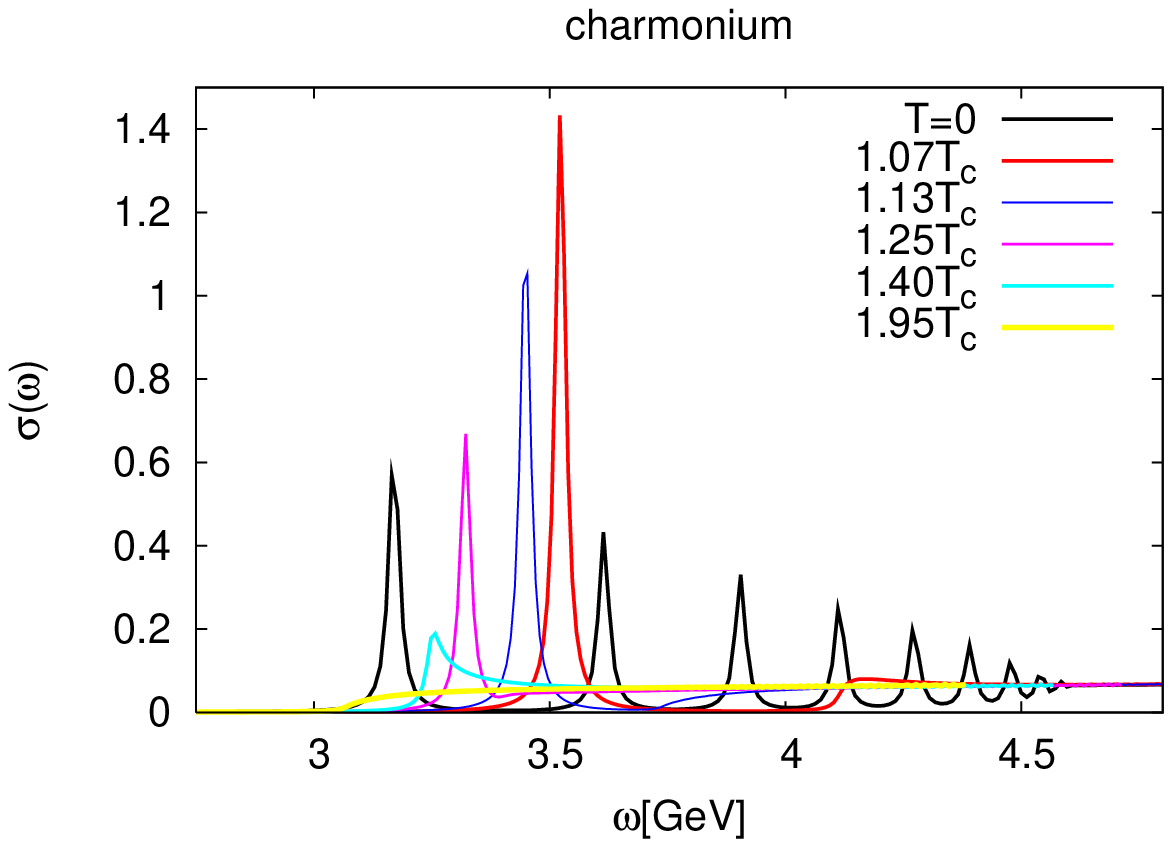,height=46mm}
\end{minipage}
\hspace*{2cm}
\begin{minipage}[htbp]{5cm}
\epsfig{file=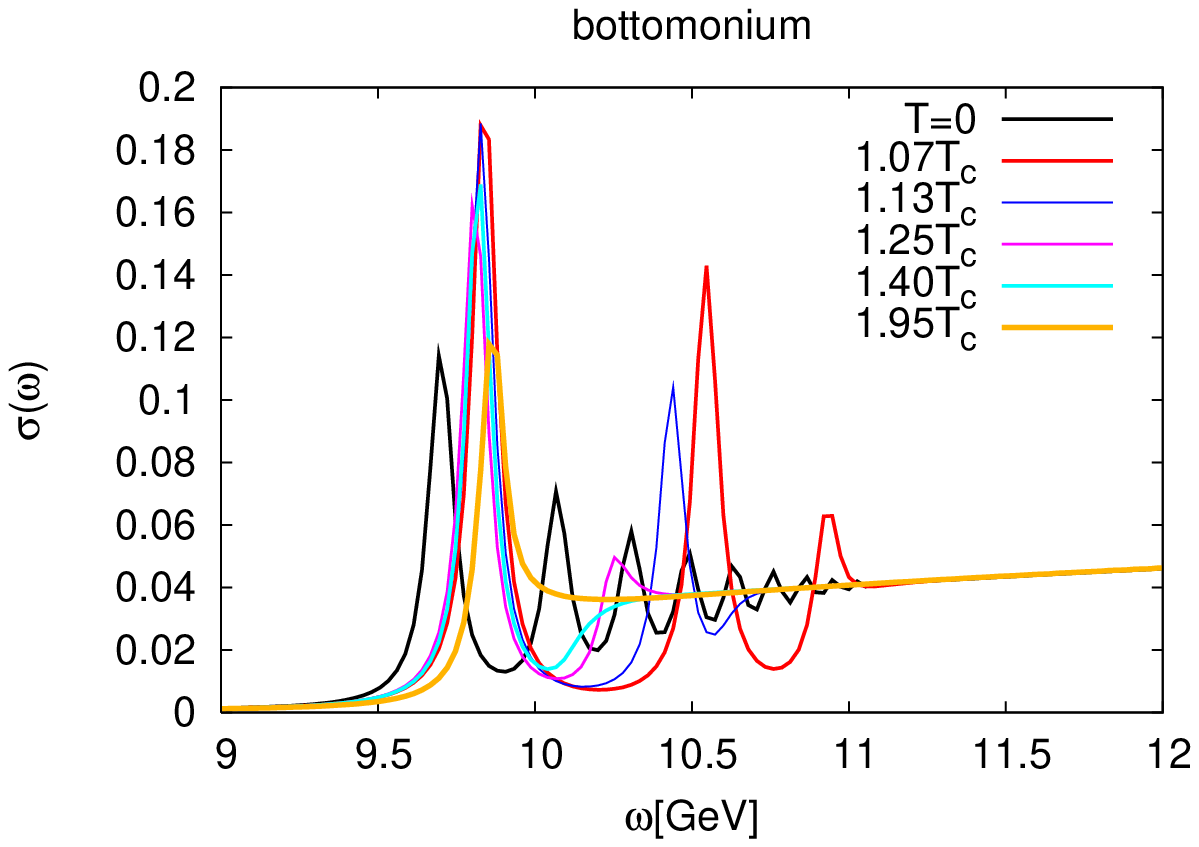,height=46mm}
\end{minipage}
\caption{Spectral function of S-wave charmonium (left panel) and bottomonium (right panel) at different temperatures obtained with the internal energy.} 
\label{fig:internal}
\end{figure}

The spectral function determined using the lattice internal energy as potential is shown in Fig.~\ref{fig:internal} for the S-wave charmonium (left panel) and bottomonium (right panel). One can clearly see that the ground states indeed survive well above $T_c$, while the higher excited states melt into the continuum. One can also identify a drastic change in the mass of the ground states. Such a few hundred MeV shift is not detected on the lattice. There is also a large increase of the amplitude above $T_c$ compared to its $T=0$ value, which is not supported by the lattice. The increase of the amplitudes is the result of the sharp rise of the internal energy above $T_c$, rise that makes the interpretation of this as potential conceptually difficult. 

\begin{figure}[htbp]
\begin{minipage}[htbp]{5cm}
\epsfig{file=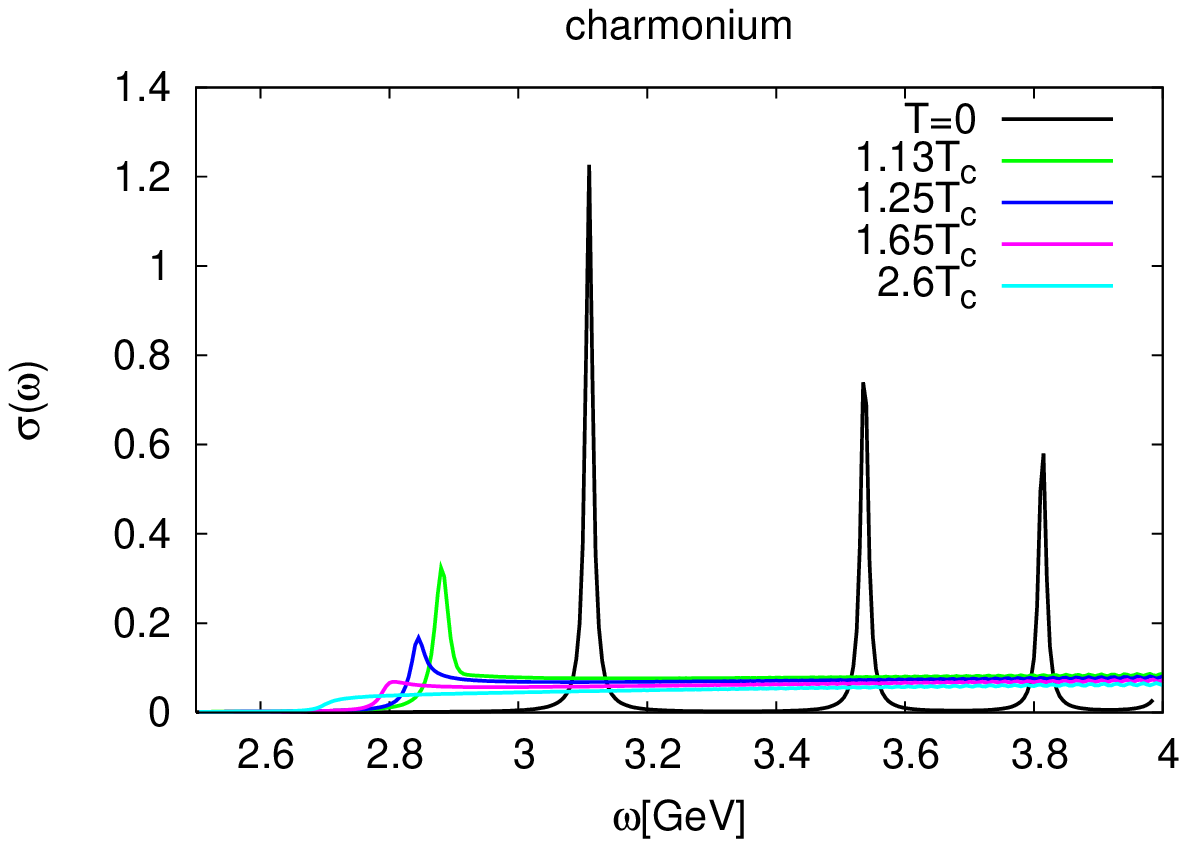,height=46mm}
\end{minipage}
\hspace*{2cm}
\begin{minipage}[htbp]{5cm}
\epsfig{file=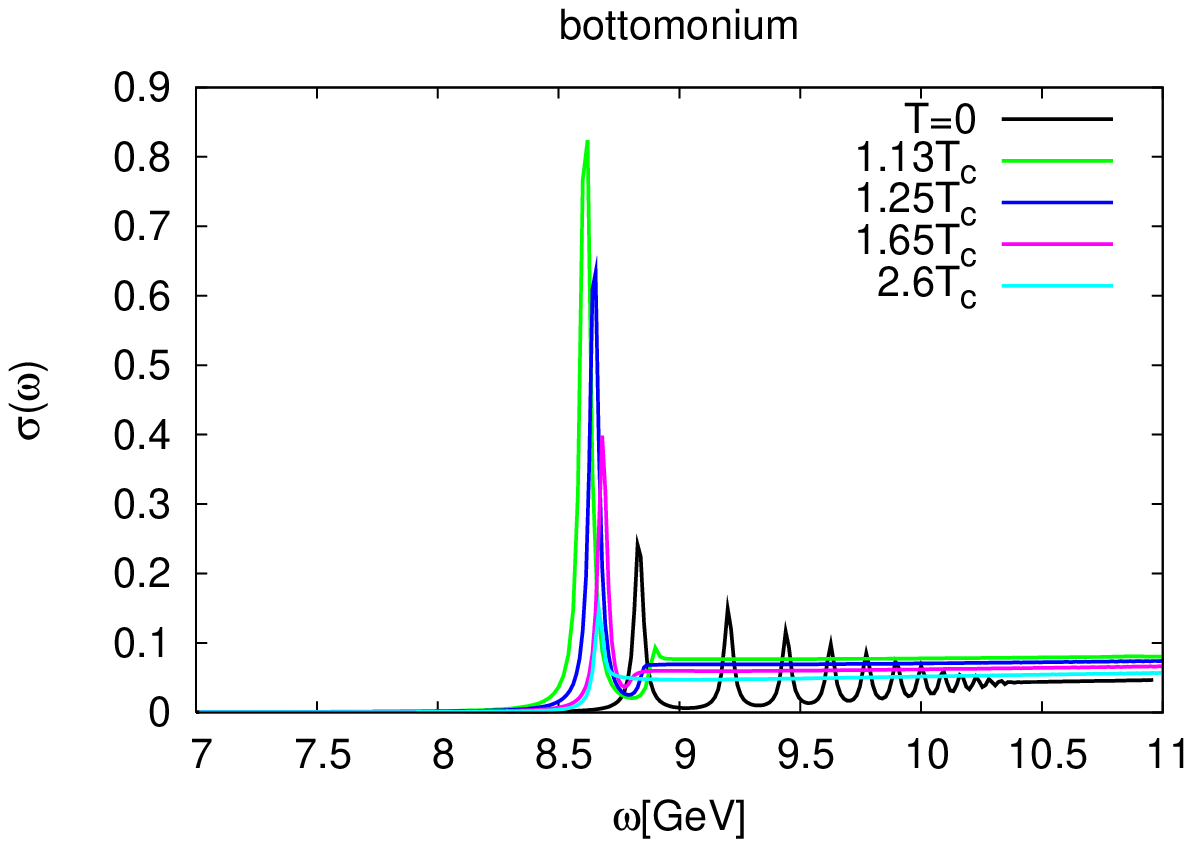,height=46mm}
\end{minipage}
\caption{Spectral function of S-wave charmonium (left panel) and bottomonium (right panel) at different temperatures obtained with Wong's potential.} 
\label{fig:wong}
\end{figure}

The spectral functions calculated with the potential suggested by Wong are presented in Fig.~\ref{fig:wong} for the S-wave charmonium (left panel) and bottomonium (right panel). The $c\bar{c}$ spectral function shows a huge decrease of the mass and amplitude of the 1S state near $T_c$, an effect not observed on the lattice. The survival of the ground state in this case is merely a threshold enhancement effect. There is a sizable decrease of the 1S bottomonium mass and an increase of its amplitude, also not seen on the lattice.  

\begin{figure}[htbp]
\begin{minipage}[htbp]{5cm}
\epsfig{file=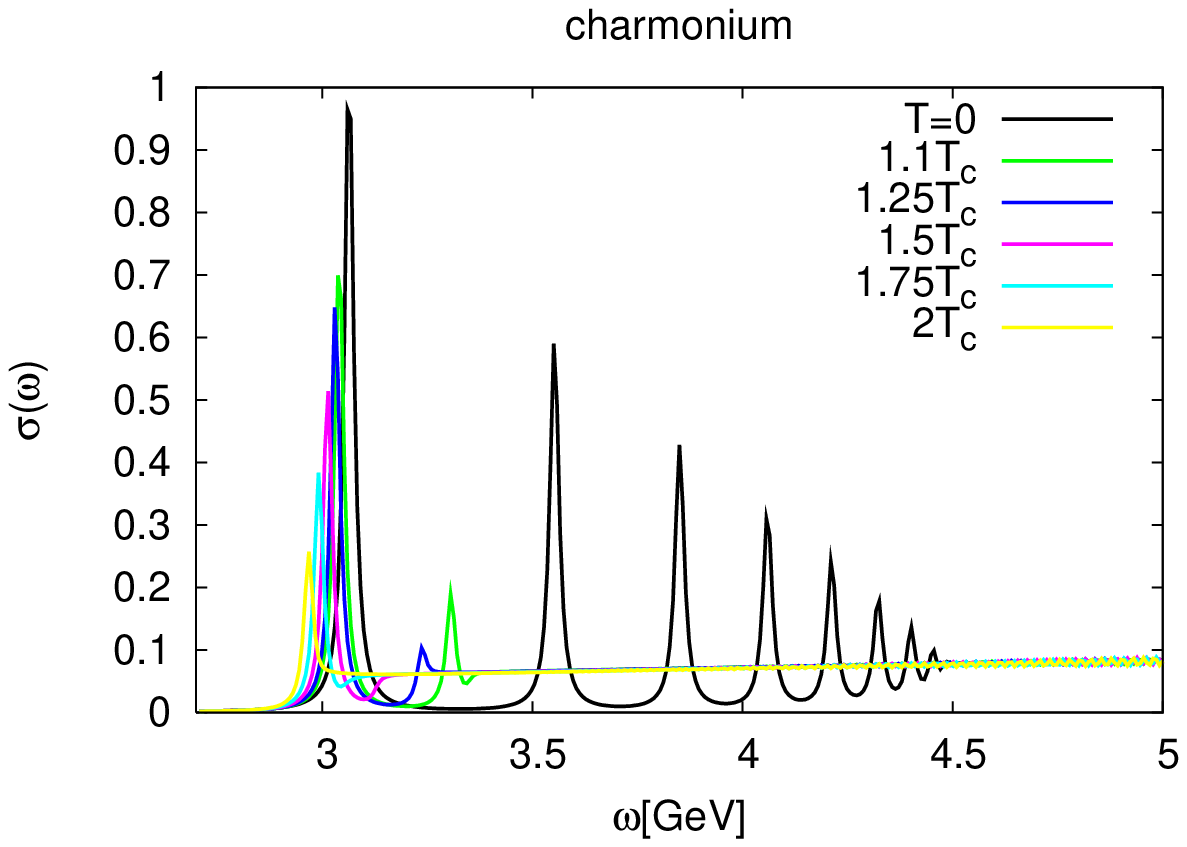,height=46mm}
\end{minipage}
\hspace*{2cm}
\begin{minipage}[htbp]{5cm}
\epsfig{file=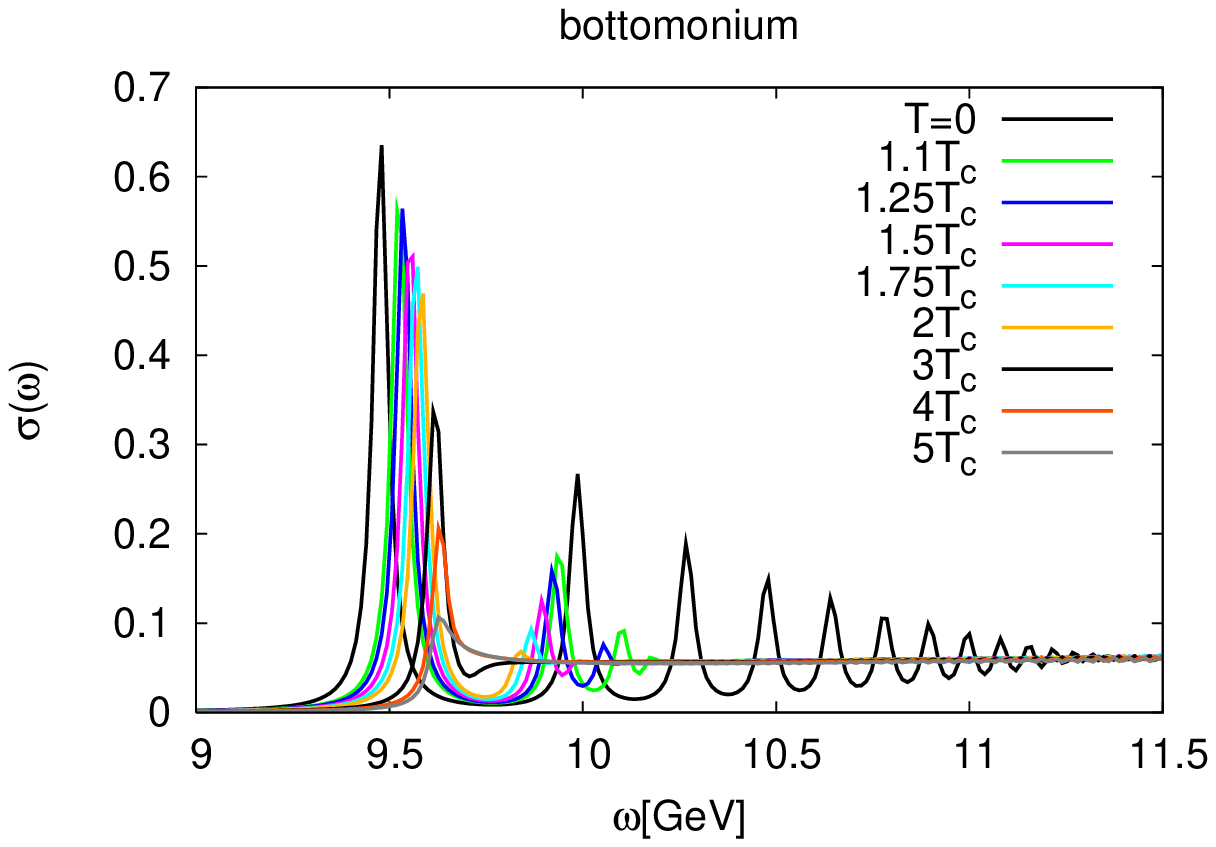,height=46mm}
\end{minipage}
\caption{Spectral function of S-wave charmonium (left panel) and bottomonium (right panel) at different temperatures obtained with the "screened" Cornell potential.} 
\label{fig:cornell}
\end{figure}

We also consider a "screened" Cornell potential as input. Here the screening length is a parameter that is related neither to Debye-screening, nor has anything to do with the free and internal energies determined on the lattice. Our preliminary finding is that we can tune this parameter such that qualitative agreement for the S-waves can be obtained. This is illustrated in Fig.~\ref{fig:cornell} for the charmonium (left panel) and bottomonium (right panel) channels. 

We conclude that even though the ground state quarkonia can survive well above $T_c$ in lattice fitted screened potentials, the spectral function is strongly modified already near $T_c$, in contrast with the findings from lattice. The temperature-dependence of the orbital spectra is ongoing investigation. Clearly, further studies are required to understand the medium modification of quarkonium spectral functions.

\end{document}